\documentclass[12pt, a4paper,superscriptaddress,groupedaddress]{revtex4}   
\usepackage{amsmath}
\usepackage{bbm}
\usepackage{graphicx}  
\usepackage{dcolumn}  
\usepackage{bm}        
\usepackage{amssymb}   
\usepackage[section]{placeins}
\usepackage{bbm}
\usepackage{float}

\newcommand{\q}[1]{``#1''}

\begin{document}

\title{A theory of interpretive clustering in free recall} 

\author{Francesco Fumarola}
\noaffiliation

\date{\today}
 
\begin{abstract}   
A stochastic model of short-term verbal memory is proposed, in which the psychological state of the subject is encoded as the instantaneous position of a particle diffusing over a semantic graph with a probabilistic structure. The model is particularly suitable for studying the dependence of free-recall observables on semantic properties of the words to be recalled. Besides predicting some well-known experimental features (contiguity effect, forward asymmetry, word-length effect), a novel prediction is obtained on the relationship between the contiguity effect and the syllabic length of words; shorter words, by way of their wider semantic range, are predicted to be characterized by stronger forward contiguity. A fresh analysis of archival data allows to confirm this prediction. 
\end{abstract}     
           
\maketitle

\section{Free Recall: Matrix Models and Graph Models}

Free-recall experiments are a key tool for the controlled investigation of episodic memory. A typical free-recall experiment takes place in two stages: during the \q{presentation stage}, subjects are shown a list of words; during the \q{memory test}, they are requested to recall them in any order. 
 
Some of the main effects reported are: 

(1) power-law scaling: the number of retrieved items scales like a power law of the number of items in the list (Murray et al., 1976); 

(2) the contiguity effect: items contiguous within the list tend to be recalled contiguously (Kahana, 1996); 

(3) forward asymmetry, i.e. the tendency to recall items in forward order (already reported in Ebbinghaus, 1913);

(4) the word length effect: lists of shorter words are recalled better than lists of longer words (Baddeley et al., 1975).

The contiguity effect and several other phenomena are now well understood by means of retrieved-context theories of episodic memory such as the Temporal Context Model of Howard and Kahana (2002). In these theories, the recovery of a memory is mediated by the recovery of its \q{temporal context}, and temporal contexts are modeled through a matrix representation that undergoes a linear evolution in time.   
  
While the effectiveness of these theories is undisputed, recently Romani et al. (2013) have introduced a somewhat different approach to the modeling of free recall, based not on matrix methods but on what we will refer to as semantic graphs.

Each node of the graph corresponds, in this description, to a different word in the list.  Retrieval is effected by a diffusive particle moving over the semantic graph, and the subject's psychological state at each moment is encoded as the current position of the particle. Whenever the particle moves onto a certain node, the word associated to that node is recalled. Diffusion is terminated when the path self-intersects.
 
This scenario was introduced as a toy version of a subtler neural-network model and was used it to compute explicitly the power-law scaling of retrieval. The power-law exponent was found to be $1/2$, which is indeed close to experimental values. This is a substantial result that may not have been as easy to obtain through more conventional theories, and it encourages further exploration of graph methods in the study of free recall. 

In this paper, a versatile graph-based model is proposed and, in its simplest form, is used to extract observable quantities. Some well-known free-recall effects are thus given a possible new explanation, while a novel effect is predicted and verified by comparison with data. 

I begin by introducing, in the next section, a more realistic family of semantic graphs, allowing for both random edges and multiple meanings, and I proceed to demonstrate that the resulting theory exhibits both the contiguity effect and forward asymmetry. I then recall some well-established results from linguistics concerning the correlation between meaning and word-length. Applying these to the diffusive-particle model yields a prediction on the correlation between word-length and the contiguity effect. 
This prediction is checked through an original analysis of archival free-recall data. Finally, I show that the underlying mechanism may also provide a semantically grounded explanation for the word-length effect. 
  
\section{The semantic graph as a random graph}
 
One assumption made in Romani et al. is that the semantic graph is complete. In the most general case, however, we have no detailed information on the structure of the semantic graph; moreover, because semantic associations are built through individual experience, they vary from subject to subject over any population. Both uncertainty and variability can be taken into account by assuming the graph's edges to be chosen at random. 
The semantic graph is then a probabilistic graph with a fixed number of nodes.

Call $P(G)$ the probability weight for a specific graph $G$. The full distribution $P(G)$ encodes the probabilistic structure of the semantic graph. The quantities we want to predict (recall probabilities) can be computed by simulating trajectories on each possible graph $G$ and averaging results over all such graphs, the average being weighed with the factor $P(G)$.
 
The theory has thus become conceptually closer to conventional retrieved-context theories.  In a non-complete graph, indeed, some pairs of nodes are more closely connected than others, and since memories are located on individual nodes, we can say of a memory that it belongs to a certain \q{region} of the graph. This is equivalent to the statement that a memory belongs to a certain \q{context} in retrieved-context theories. The semantic graph may be seen, therefore, as describing the structure of contextual states (Kahana and Howard, 2002), which is a probabilistic structure both because of our ignorance of its details and because of its variability across a population. 

A graph model of free recall involves this probabilistic structure in a direct and explicit way, and may therefore become helpful as part of an endeavor to elucidate the semantic graph empirically. Connections between various semantic contexts are fully encoded in the distribution $P(G)$; if we compute the recall probabilities for various choices of the distribution $P(G)$ and compare them with the experimental values, the \q{true} structure of the semantic graph will be the choice of $P(G)$ that yields the best agreement with the data.

\begin{figure}[h!]
\label{1} 
\includegraphics[width= .8 \textwidth]{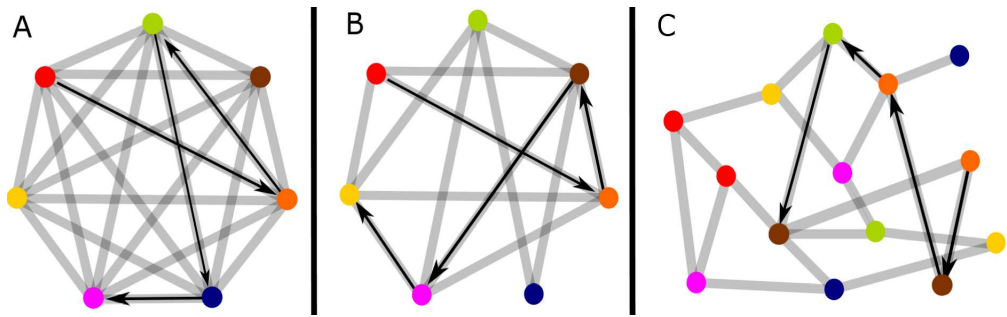}
\caption{Panel A: particle diffusing through a complete graph as in the model of Romani et al., 2013. Panel B: diffusion on a non-complete graph. Panel C: diffusion on a non-complete graph with the inclusion of polysemy; in this particular example, each word has two semantic nuances, or \q{meanings}, represented by as many nodes. Nodes of the same color represent different meanings of the same word.}
\centering 
\end{figure} 

In this paper, we try out what is arguably the simplest trial distribution $P(G)$, corresponding to a $G(N,p)$ Erd\H{o}s-R\'enyi  model with $p=1 - \alpha$. In other words, the edges of the complete graph are removed independently on each other, each having a probability $\alpha$ of being removed. If $\alpha=0$, the semantic graph is complete; for arbitrary values of $\alpha$, the probability associated to a given graph with $n$ edges is  $P(k) = (1-\alpha)^n  \alpha^{\frac{N (N-1))}{2}  -  n}$ . 
 
\section{Introducing Polysemy}
   
Before computing measurable quantities, i.e. recall probabilities, we must notice a second limitation to the model used by Romani et al.. The graph they employ represents every word in the vocabulary as a single node. On the other hand, fMRI measurements have convincingly shown that the neural response to free-recall tests exhibits a strong statistical dependence on the semantic variability of words (Musz and Thompson-Schill, 2015).   

In linguistics, the degree of dependence of a word's meaning on context is called polysemy (Nerlich et al., 2003). Of course, since meaning is inevitably affected by context, no word is perfectly \q{monosemic} (i.e. having a single nuance of meaning), but a word with comparatively little semantic variability is called \q{oligosemic} (Fernando, 1996). To graft polysemy into the graph model, we must identify the nodes of the semantic graph with \q{meanings} (or semantic nuances) rather than with words, allowing each word to label multiple nodes. 
 
A word $W$ will then have a degree of polysemy $k(W)$, defined as the number of nodes corresponding to word $W$. In the simplest scenario the degree of polysemy will have a constant value $K$, the same for all words (Fig. 1C).

If the semantic graph is complete, each node will be linked to $K-1$ nodes corresponding to the same word, and to $K$ nodes corresponding to every other word in the vocabulary. If the semantic graph is the Erd\H{o}s-R\'enyi  graph $G(KN, 1-\alpha)$, with nodes labeled by the $N$ words, a node corresponding to any given word will be linked on average to $(1 - \alpha) K$ nodes corresponding to every other word, a well as $(1 - \alpha) × (K-1)$ same-word nodes.
 
Given that each word corresponds to multiple nodes, one question arises concerning the retrieval process. Will a word be recalled when the diffusive particle touches \textit{any} of the nodes corresponding to it? Or are memories encoded each in a given node, even when the word to be recalled has multiple meanings? 

The literature on context-retrieved theories suggests that the latter option holds true. Indeed, we know that memories are anchored to the contextual region where they have been created during the presentation of the list (Howard and Kahana, 2002). If a word has multiple meanings, its recall will require retrieving the specific meaning that was attributed to that word during presentation.

In order to know which node corresponds to a given memory, we need to formalize the dynamics during presentation, which can be simply modeled as another diffusive process on the graph. At every instant during the presentation stage, the diffusive particle lies on a definite node; once a word is presented, the particle diffuses until it \q{recognizes} that word, i.e. until it stumbles on one of the nodes corresponding to it. 

This process has an interpretive function: the system \textit{interprets} each word through the meaning of that word on which the diffusing particle stumbles first, and that particular node becomes the location of the memory corresponding to the word.    
 
Notice that this recognition may never occur, as the Erd\H{o}s-R\'enyi graph has a finite probability of being disconnected; if no available path leads from the current position of the particle to any of the word's nodes, the particle will be allowed to jump on to a node randomly chosen amongst them.

This interpretive process takes place for each word in succession: once a word has been interpreted, the next word in the list is presented, and diffusion goes on. Thus memories are created. 

\begin{figure}[h!]
\label{2} 
\includegraphics[width= \textwidth]{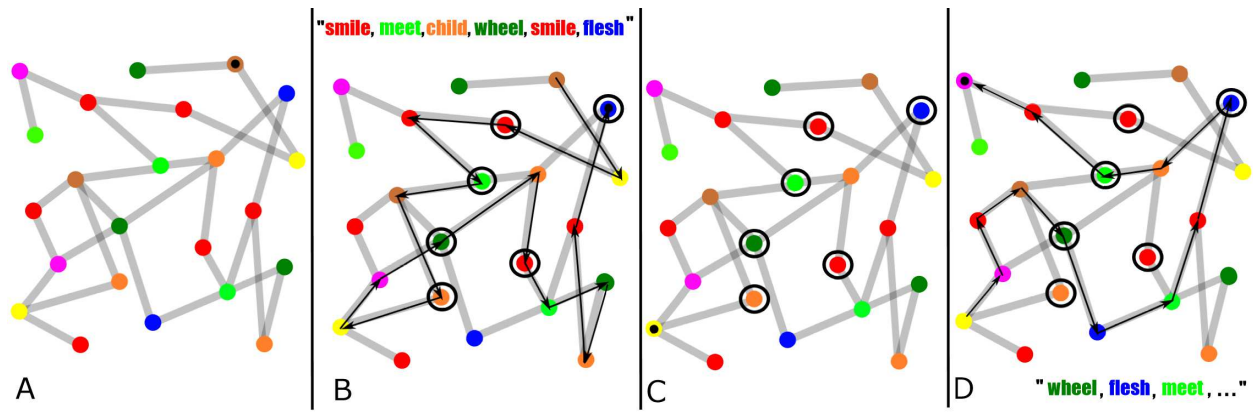} 
\caption{Diffusive-particle model of a free-recall experiment. Panel A: a semantic graph, shown with a specific choice of its edge structure among the many such structures over which final results must be averaged; meanings corresponding to the same word are shown in the same color; the current position of the particle is indicated by a black dot. Panel B: presentation stage; each time a new word is presented, the particle keeps diffusing until it lands onto any of the meanings described by that word; the resulting trajectory is shown as a sequence of arrows. Panel C: the nodes where meaning has been found during presentation have become transient memories (circled nodes); in the interval between presentation and memory test, the diffusive particle's position is reset to a random point indicated by the black dot. Panel D: during the memory test, a new diffusive process takes place, similar to the one described in Romani et al., 2013. The diffusive particle has to locate the circled nodes for the corresponding words to be recalled.
}
\centering 
\end{figure}

The model includes therefore two diffusive trajectories: one effecting the interpretation of words and one effecting their retrieval (as in Fig. 2). These two trajectories, in fact, are meant to model processes rooted in different cognitive abilities, so it would be more correct to speak of two different \q{particles}, one employed for interpretation and one for retrieval. In practice, we only need to be cognizant that the two diffusive trajectories may develop at different speeds. 

\section{Contiguity Effect and Forward Asymmetry}
  
Let us call \q{transition} any sequence of two consecutively recalled word.  Obviously, neither the first word recalled in the test-stage of an experimental trial nor a word recalled right after an intrusion are part of a transition. We will call \q{lag} the difference between the serial positions of two words in a given transition; for example, if the $5$th word in the list is recalled right after the $8$th, the corresponding lag is $L= -3$. 

In addition, let us call $p(L)$ the lag probability distribution, i.e. the probability that an arbitrary transition will have a lag $L$. In terms of this distribution, the contiguity effect may be described as the experimental fact that $p(L)$ is a decreasing function of the absolute value $|L|$ for $|L| \geq 1$. Forward asymmetry, on the other hand, is the fact that $\sum_{L>0} p(L) >\sum_{L<0} p(L)$ , i.e. lags are more often positive than negative, meaning that forward transitions are preferred; as we will see, this fact is due almost entirely to the contribution from contiguous transitions ($L=±1$).   

To check for these phenomena, we proceed to simulate the diffusive-particle model for a large number of different graph structures. These graphs corresponds to fixed parameter values $N$ and $K$ (number of nodes and their associated polysemy) but to different combinations of edges. For each graph, we have to mimic the presentation of a large number of different lists. We then let the retrieval process take place, and compute the probabilities by averaging over trials. Results for different graphs must be summed up with the weight factor $P(n)$ mentioned above.

The observed frequency of transitions with a given lag is plotted in Fig. 3A, for various choices of the vocabulary size  $N$, the polysemy level $K$, and the semantic disconnectedness $\alpha$. To obtain this curve, pseudodata have been generated on a large number of free-recall trials, and the number $n(L)$ of recall events with lag $L$ has been divided by the total number of recall events, so as to yield an estimate of the lag probability $p(L)$. 
 
The main features of the curve, as can be seen, are invariant for various combinations of parameter values. As we are not considering repetitions, by construction the curve vanishes at $L=0$; there are two maxima at $L= ±1$; and the transition probability is a decreasing function of $|L|$, the absolute value of the lag. This amounts to the statement that the contiguity effect is predicted by this model.  

Moreover, the curve is not symmetric around $L=0$: the forward branch sums up to a larger cumulative, although it lies higher up only insofar the peak at $L=1$ is concerned. I will refer to this peak as the \q{sequential} peak, and to forward contiguous transitions as \q{sequential transitions}. The sequential peak is always considerably higher than the backward contiguous peak -- a phenomenon widely documented in experiments (see Kahana, 2012).  

To offer an example of how these features emerge in empirical results, panel 3B displays the curve of transition frequencies for data from PEERS (Penn Electrophysiology of Encoding and Retrieval Study), a large study conducted at the University of Pennsylvania. The data are those described in Lohnas et al. (2015), summing up to a total of $7360$ free-recall trials on $92$ subjects, all performed with lists of $16$ words. Intrusions have been discarded from these data, and no availability correction has been introduced; repetitions, which are comparatively rare, have been counted in under the lag $L=0$. 

In the database concerning each subject, transition events with the same lag have been grouped, counted, and normalized by the total number of transition events to yield the subject's curve of transition frequencies. The averages of these curves over all subjects and the standard deviations of the corresponding distributions are shown respectively as the solid curve and the error bars of Fig. 3B. 

The empirical curve thus obtained and the curves obtained from simulations are not identical. Nonetheless, the features we have outlined above are prominent in both. In particular, the difference between the backward and the forward branch of the curve is concentrated in both cases at contiguous transitions, and the maximum at $L=1$ is always the global maximum of the distribution. This is a feature nontrivially displayed by the model, and the mechanism behind it will become clearer in the next sections. 

\begin{figure}[h!]
\label{3} 
\includegraphics[width= \textwidth]{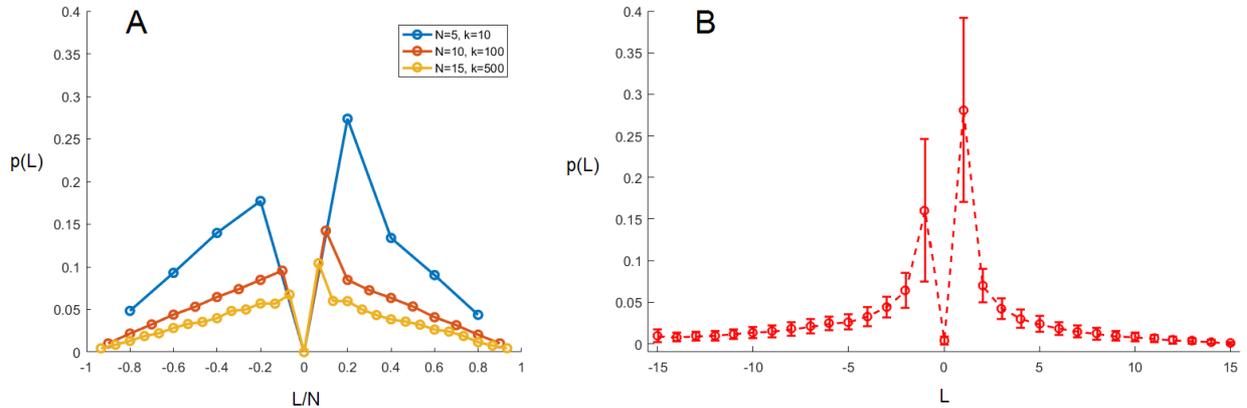}
\caption{Panel A: Results of simulations on the diffusive-particle model for three choices of the vocabulary size N and polysemy K (see legend) and for $\alpha= 1 - 1/K$. Lists presented to the model were permutations of the full vocabulary. The y-axis shows transition frequencies, the x-axis the serial-position lag normalized by the size of the lists. Panel B: Transition frequency as a function of lag, as computed from PEERS data.
}
\centering 
\end{figure}

\section{Polysemy and Word Length}
   
In the previous section, we simulated the model under the assumption that all words had the same degree $K$ of polysemy, i.e. the same number of \q{semantic nuances}. We may now wonder how the recall probability of a word varies as a function of the word's degree of polysemy. 
 
Polysemy is, unfortunately, a somewhat elusive variable and rather subtle to measure  (Nerlich et al., 2003). Consider for instance the two words \q{lion} and \q{lioness} (a classic example); does the meaning of \q{lioness} vary with context? Surely less than the meaning of \q{lion} because, aside from finer distinctions, the word \q{lion} has at least two potential meanings (a male lion, or a lion of unspecified gender), while \q{lioness} has, by comparison, just one (a female lion). Nonetheless, a typical dictionary would only mention gender in connection with \q{lioness} and not provide distinct definitions for the two meanings of the word \q{lion}.  
  
Linguists have been studying this type of problem in depth for decades (Greenberg, 1966; Jakobson, 1987). One of their most useful conclusions is that the syllabic length of words may be employed as a reliable, and easily measurable, statistical indicator for oligosemy. Said otherwise, longer words have proven to be robustly less polysemic than shorter ones, and (as in Rensinghoff and Nemcov\'a, 2010) a Waring distribution seems to fit best this dependence. For numerical details on the correlation, see the statistical studies in the literature, in particular Zipf (1949), Guiter (1974), Sambor (1984), Rothe (1994). 
 
I will measure the length of words through their number of syllables; hereafter, by \q{word-length} I will always mean the number of syllables in a word.  In the experiments of Lohnas et al. (2015), whose data I have used above, word lists were assembled from a pool consisting of 1638 words with up to 6 syllables. However, only four 5-syllable words are present, and a single 6-syllable word (\q{encyclopedia}); hence, the statistics for these two lengths may not be representative. 
  
An interesting feature that emerges from these data concerns the \q{sequential} peak of the lag probability distribution (the forward contiguous transition frequency). Suppose that the distribution is computed only over transitions to words of syllabic length $M$, so that it can be written as $p_M (L)$. It appears that the height of the sequential peak, $p_M (+1)$, exhibits a nontrivial dependence on the length $M$ of the word recalled, i.e. the probability of sequential recall varies significantly over words of different lengths. 
  
The relative frequency of sequential transitions can be estimated from the data in at least two separate ways, through a word-by-word statistics or a subject-by-subject statistics. The results from both approaches are shown in Fig. 4. 
  
Panel A of the figure shows results obtained by regarding every transition as an independent event. Call $n(S,W,L)$ the number of observed transitions to word $W$ with lag $L$ in trials on subject $S$. In Fig. 4A, each blue dot shows the value of the ratio 
$R(W) = \frac{\sum_S n(S,W,1)}{\sum_S \sum_L n(S,W,L)}$
computed for a particular word W. The histogram of this ratio over all words with the same length has been plotted vertically for each word-length $M$ (black curves). Red circles show the means of these values over all words with $M$ syllables: $m_1 (M)=  \frac{1}{| V(M)|} \sum_{W \in V(M) } R(W)$, where $V(M)$ is the set of all words with $M$ syllables used in the database, and $|V(M)|$ their number. 
 
The trend of the resulting curve is decreasing. Extracting the correlation coefficient yields $r= -0.12$, with a negligible p-value $p<10^{-5}$. This signifies that the longer a word, the smaller its chance of being recalled through a forward contiguous transition. 
 
While this is an intriguing result, it relies of course on the assumption that all transition events could be treated independently. On the other hand, transition events within the same trial are statistically correlated, and the same may be true for transition events within different trials performed on the same subject. 

In panel B, a different analysis is displayed. Instead of computing the recall statistics for each individual word, we characterize every transition event solely by the length of the word of arrival. Information on the particular word involved is ignored, i.e. assumed to average out.
 
For each subject $S$ and word-length $M$, I have computed the ratio 
$R(S,M)= \frac{\sum_{W \in V(M)} n(S,W,1)}{\sum_L \sum_{W \in V(M)} n(S,W,L)}$, and values of these ratio as shown as blue dots in Fig. 4B.; histograms of these quantities are again shown in black; the mean values 
$m_2 (M)=  \frac{1}{N_s }  \sum_S R(S,M)$  (where $N_s$ is the number of subjects) are shown as red circles. 
 
If the normalization factors depended solely on word length, i.e. in the case where $\sum_L n(S,W,L)=n(M)$ for all $S$ and all  $W \in V(M)$, we would have $m_1 (M)= m_2 (M)=   \frac{\sum_{S, W \in V(M)} n(S,W,1)}{|V(M)| N_s n(M)}$ for all $M$. This is the case, in particular, if the samples are identical over all words and subjects, which is of course not true in any realistic dataset. Nonetheless, the mean values we have obtained from the subject-by-subject statistics are fairly close to those obtained in the word-by-word statistics (Fig. 4B). 
 
Moreover, we find once again that the mean probabilities for sequential transitions are monotonously decreasing as functions of word length. As for the correlation coefficient, it is also close to the value found above: $r = - 0.11$. The $p$-value is higher, but still low enough to enable our correlation hypothesis ($p= 0.01$). All this provides substantial evidence that sequential transitions (with lag $L =+1$) are indeed better favored for shorter words. 

We should also report that no significant correlation between transition probabilities and word-length has been found for transitions with lags other than $L= +1$. For example, if the foregoing analysis is repeated for backward contiguous transitions, to estimate the dependence of $p_M (-1)$ on the word-length $M$, a p-value of the order of p $\sim$ 0.2 is obtained both from the word-by-word and from the subject-by-subject statistics -- too high for the correlation to be considered relevant. We must conclude that the effect we are describing arises from mechanisms that concern exclusively sequential transitions.     
 
Surely several explanations may be conjectured, claiming (as we will do) that the effect is rooted in polysemy means essentially claiming that it is contextual in nature; by definition, a word is the more polysemic the larger the semantic context it covers. If that is true, we expect that a retrieved-context model of free recall should be suitable to predicting the effect. The model introduced in the previous sections is particularly handy for a polysemy-centered discussion. As I will show in the next section, it also provides a particularly simple explanation for the phenomenon we just singled out. 
 
\begin{figure}[h!]
\label{4} 
\includegraphics[width= \textwidth]{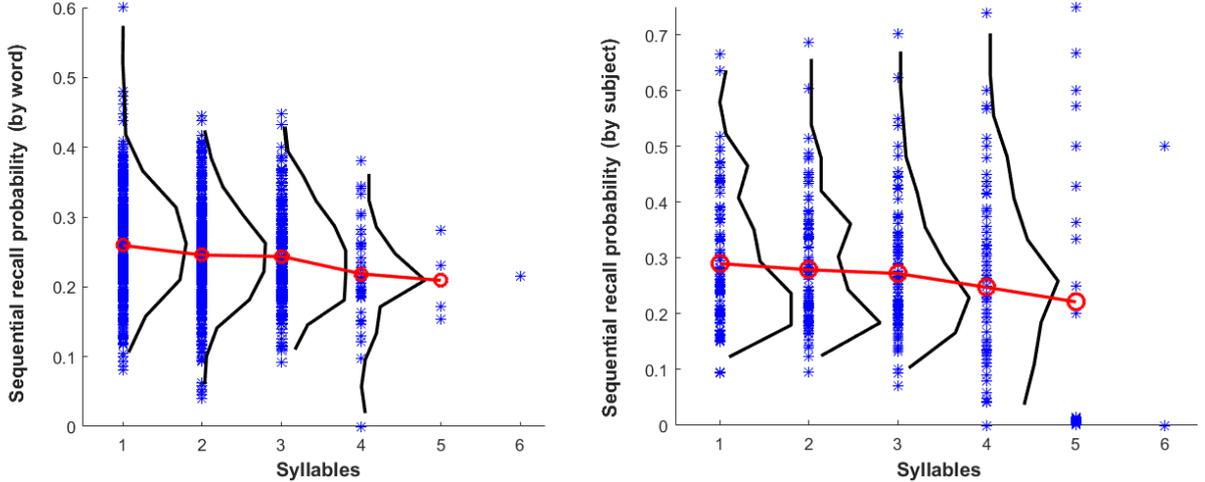}
\caption{Panel A: Probability that a word, if recalled, will be recalled sequentially, computed from PEERS data by regarding all recall events as independent. Each blue dot corresponds to a different word; for example, the high-lying one-syllable outlier is the word \q{belt}. The black curves are histograms of these probabilities over all words of a given length, as indicated on the x-axis, and red circles indicate their means. Panel B: Probability that an individual subject will recall a word of a given length sequentially, obtained from PEERS data by regarding all words of the same length as equivalent. Each blue dot corresponds to a different subject; points overlapping at zero have been jittered for display; histograms over all words of the same length are shown as black curves, their means as red circles. 
} 
\centering 
\end{figure}

\section{Interpretive Clustering}
    
Let us consider the above semantic graph in the case where the polysemy $k(W)$ of word $W$ varies over different words, i.e. word $W$ has $k(W)$ semantic nuances.  

The quantity we would like to calculate is the lag probability distribution $p_k (L)$, i.e. the conditional probability that a word with k semantic nuances, if recalled, will be recalled through a transition with lag $L$. If the effect we observed in the experimental data is indeed due to polysemy, we should expect the sequential transition probability $p_k (1)$ to be enhanced for more polysemic words. Moreover, because of the normalization constraint $p_{k} (1)+ \sum_{L \neq 1} p_k (L)=1$, this entails that the probability $p_k (L)$ for any $L \neq 1$ should be comparatively suppressed with more polysemic words.  

Fig. 5 shows the results of simulations on a semantic graph with $\alpha = 0.9$. The lists presented to the system were permutations of the whole vocabulary; the conditional probability $p_W (L)$ that a word $W$, if recalled, will be recalled with a lag $L$, has been averaged over all words with the same degree of polysemy $k(W)$ and the means are displayed as bar plots of different colors. 
Panels A, B, and C refer to results for a vocabulary of 2N words of which N are monosemic (i.e. have one meaning) and the remaining N words are disemic (i.e. have two meanings). The values of N are respectively 2,3,4, as shown over the plots, and all three yield qualitatively identical plots.  

The most conspicuous feature of these plots is the sequential peak exhibited by the disemic word as opposed to the monosemic one. The sequential recall probability $p_k(L=1)$ is hence a sharply increasing function of polysemy (or a decreasing function of word length, just as found in the data). On the other hand the lag probability distribution for each word-type is normalized, so this gap should be made up for in nonsequential transitions. Indeed, we observe that nonsequential transitions are slightly more frequent for the monosemic words than for the disemic one, the difference at $L=1$ being re-distributed over all nonsequential values of the lag.

\begin{figure}[h!]
\label{fig5} 
\includegraphics[width= \textwidth]{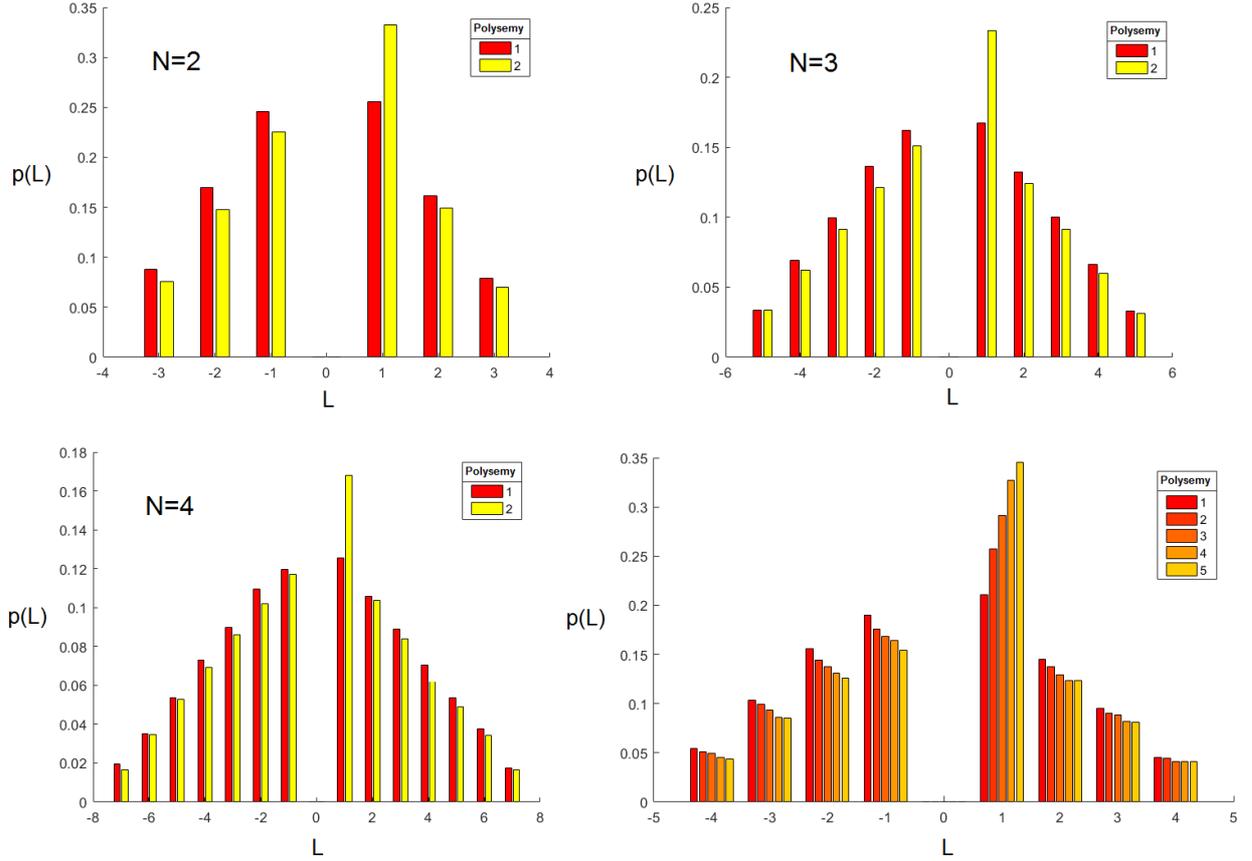}
\caption{Lag probability distribution $p(L)$ from simulations where the lists presented for recall are permutation of the vocabulary, and the semantic graphs correspond to $\alpha=0.9$. Different bar colors refer to different degrees of polysemy $k$, shown in the legends. Panels A, B, C: results for a vocabulary of $2N$ words of which $N$ are monosemic (i.e. have one meaning) and $N$ are disemic (two meanings); the values of $N$ are shown over the plots. Panel D: results for a five-word vocabulary in which each word has a different degree of polysemy (from $k=1$ to $k=5$). }
\centering 
\end{figure}
    
We may ask whether the correlation between sequentiality and polysemy holds true also for words with more than two meanings. As an illustration, panel C of Fig. 5 displays results of simulations for a vocabulary of $5$ words, one for each degree of polysemy between $k=1$ and $k=5$. The overall picture that emerges is a straightforwarad extension of what has been found in the case of only two word-types: again, the sequential probability $p_k (1)$ is a sharply increasing function of a word's degree of polysemy $k$; again, all other values of $p_k (L)$ are faintly decreasing functions of polysemy.  

We conclude that the positive correlation between sequential recall and polysemy is a feature robustly displayed by this model. The more meanings a word has, the more easily it is recalled in the order in which it was presented. The question is now why this happens – i.e., what is the ubiquitous mechanism lying at the root of this relationship. 

To answer this question, we recall that, by introducing a degree of disconnectedness in the semantic graph, we have endowed it with a nontrivial topology, in which some meanings are closer to each other while others lie further apart. A possible way to measure the distance between any two nodes on a graph is, for instance, by the length of the shortest path connecting them, or by the mean time it takes to diffuse from one to the other. 
 
It is in this spirit that one should regard Fig. 6, where the distance between any two nodes represents the distance between them (i.e. length of the shortest path, or time for first passage) within a wider semantic graph. Of the graph, only a few nodes are shown -- those corresponding to three words (Red, Green, and Blue). 

Green and Blue are monosemic words; Red is monosemic in the semantic graph of panels A-B-C of Fig. 6, and polysemic in the semantic graph of panels D-E-F (having two meanings). The arrays of colored squares over the drawings in panels B-C and E-F represent lists of words presented to the system for a free-recall trial. 

In panels B and C, since all words are monosemic, memories of each word can only be created at a fixed node, and a different order of presentation does not generate different memories. Hence, Red has the same probability of being recalled after Green or after Blue. 

In panels E-F, on the contrary, the memory created by presenting the word \q{Red} tends to lie close to the memory created by the word that precedes it. This happens because Red is polysemic, so the system can \q{choose} a meaning for it. If the graph is not too disconnected, the diffusive process that interprets words is continuous (i.e. jumps are rare), so a meaning close to the current position of the particle will be more likely to be hit first.  

In panel E, therefore, Red is more likely to be recalled after Blue than after Green, while in panel F, Red is more likely to be recalled after Green than after Blue. In both cases, Red is most likely to be recalled right after the word that precedes it in the list. Thus, the polysemy of Red makes it more likely to be recalled sequentially.  

We will refer to this phenomenon as \q{interpretive clustering}: among the multiple meanings of an input, the cognitive system selects the one that fits best the content of the ongoing discourse. The more polysemic a word, the more numerous the meanings the system can choose from; hence, the more likely it is to find a meaning close by. This will logically translate, during the test stage, into an enhanced probability for sequential recall. 

\begin{figure}[h!]
\label{fig6} 
\includegraphics[width= .75 \textwidth]{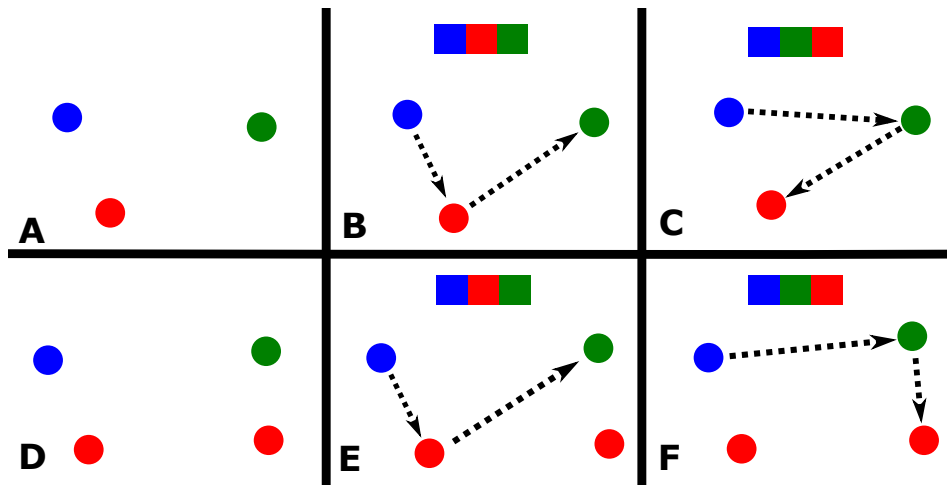}
\caption{Nodes corresponding to three words (Red, Green and Blue) within a wider semantic graph; distances on the page are meant to represent roughly shortest-path distances within the graph. Green and Blue are monosemic words; Red is monosemic in the semantic graph of panels A-B-C, polysemic in the semantic graph of panels D-E-F, with two meanings. The arrays of colored squares over figures B-C and E-F represent word-lists presented to the system. Dotted arrows depict diffusive motion through the semantic graph during presentation.
}
\centering 
\end{figure}

\section{Optimal Sequentiality}
 
It is well-known in the literature (Farrell, 2012), that a word-list presented for a free-recall test is effectively divided by the memory-storage process into \q{sequential chunks}, sections that tend to be recalled in sequential order. These chunks and their optimal length have been subjected to extensive studies (see e.g. Cowan, 2001). 

Indeed, if the peak at $p_W(L=+1)$ is large for a series of consecutive words, these are likely to be recalled in the order in which they were presented. With high probability, the peak will \q{guide} the recall process through a full sequential chunk, and the last word of the chunk will be the first where the peak is suppressed; at that point, the recall process becomes fully associative, and free association decides which chunk will be recalled next. 

The probability value $p(+1)$ approaches unity only for rare subjects (Healey et al., 2014); the peak value is, on average, of the order of $0.3$ (Fig. 3). Hence, even where information has been stored the most sequentially, the retrieval process has a finite probability of occurring in non-chronological orders.  

The sequential peak, nonetheless, is regularly the global maximum of the probability distribution $p(L)$, and this fact makes it possible to retrieve the chronological events with arbitrary accuracy, as one could easily argue in terms of diffusion. 

If the chronological ordering is the most probable, a diffusive process has indeed a particularly simple way of singling it out with arbitrary accuracy; it is sufficient to re-explore the same contextual area a large number of times, and to choose the ordering of memories that has been experienced most often during this re-exploration. The more strictly sequential the memory storage is (i.e. the larger $p(1)$ ), the less time it will take to perform the iterative sampling needed to establish a chronology with arbitrary accuracy.

It can then be conjectured that the value of $p(1)$ is optimized to compromise between two conflicting goals:  (1) to allow for a fast-enough iterative sampling -- as described -- and (2) to keep the memories available nonetheless for use by free association.

If the sequential peak is too low, the number of iterations needed to find the most probable ordering will become large, and the iterative sampling procedure slow; it may be impractical to devote more than a fraction of a second to ordering any sequence of past events.

If, on the contrary, the sequential peak is too high, associative retrieval of a given memory will be blocked, as follows from the normalization of probabilities; if we can only arrive at a memory from its chronological precedent, it cannot be accessed other than chronologically. It is consequently not available for associative tasks and becomes useless for most cognitive purposes.
  
Thus, sequentiality and retrievability are in conflict and a trade-off between the two requirements may be necessary. A memory must stay available for associative reasoning and yet its chronology needs to be trackable through iterative sampling. From these two constraints, the optimal value of the $p( +1)$ may be determined. 

This optimization process can further depend on the particular memory involved. In other words, what has been referred to as \q{chunking} may be a process based partly on a distinction between memories that need chronological storage and memories that do not. 

The suggestion of this paper is that polysemy may be one of the criteria for this distinction. As long as words with adaptable meanings are being presented, the system may keep grafting them easily into the ongoing semantic chunk. But when a word with a highly specific meaning appears, there are few chances that the current discourse may accommodate it logically. Hence a rift in the storage process would have to be introduced – and a new chunk will begin.

This may also be understood as implementing a principle of least effort (Zipf, 1949). Polysemy compels the receiver of verbal input to choose one of many possible understandings, and that is done on the basis of the chronology of the input. Thus, chronology is an important part of polysemic communication, but is less essential when the words being used are oligosemic. Memorizing the chronology of an input is, on the other hand, less useful when it does not play a role in determining its meaning. Hence, chronological information related to oligosemic words can be more safely discarded. 

\section{Word Length Effect}
 
The empirical fact that lists of shorter words are easier to recall (word-length effect) is one of the early findings in the history of free recall (Baddeley et al., 1975). Theories of this effect may be classified as being either item-based or list-based, i.e. they impute the effect either to an individual property of words or to a global property of a list.  

Recently, item-based theories have been cast doubt upon by new experiments; in particular, it appears that in experiments with mixed lists (composed of words of various lengths) shorter words are not always easier to recall (Hulme et al., 2004; Xu et al., 2009; Katkov et al., 2014). This suggests that the word-length effect in pure lists may exist \textit{not} because shorter words are more distinctive, but in spite of the fact that they are not, strongly pointing toward a list-based explanation for the effect. 

In list-based theories, on the other hand, the global property on which the effect is made to depend is most frequently the total duration of the list (Baddeley, 2007). But this explanation has now also been called into question. Neath et al. (2003) have shown that with words having the same number of syllables but diﬀerent pronunciation times, no unambiguous word length effect arises. This suggests that the effect may depend on the number of syllables and not on the time it takes to pronounce them (Campoy, 2008). A review of the debate may be found in Jalbert et al., 2011, where it was argued that \q{the word-length effect may be better explained by the differences in linguistic and lexical properties of short and long words rather than by length per se}.
 
Could this elusive linguistic property be just polysemy? This hypothesis seems not to have yet been explored, and the diffusive-particle model may help to test it. To do so, I have simulated the model by presenting lists that contain words with a fixed degree of a polysemy, while keeping the semantic-graph structure unchanged. 
 
\begin{figure}[h!]
\label{fig7} 
\includegraphics[width= .75 \textwidth]{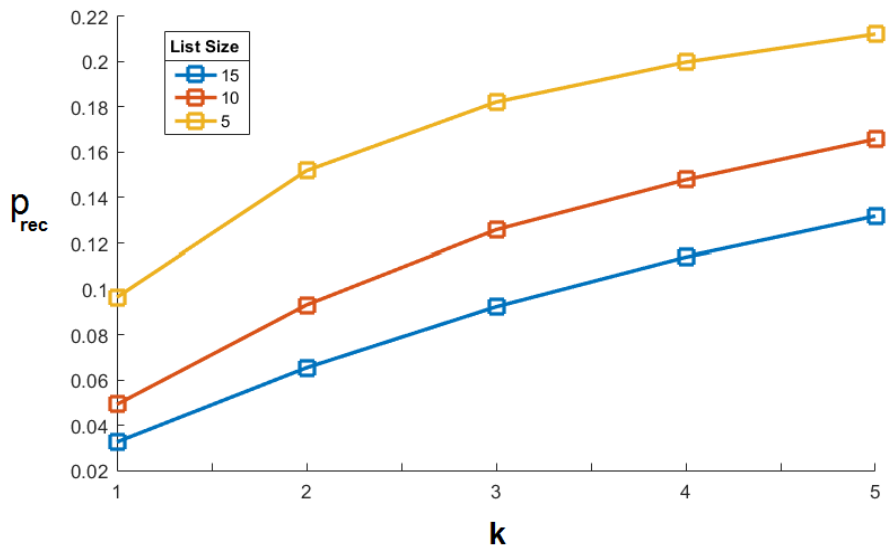}
\caption{Mean recall probability in the diffusive-particle model. The semantic graph employed for the simulations contains a vocabulary of ten words, two for each degree of polysemy between $k=1$ and $k=5$, while $\alpha  = 0.7$. The word-length effect was checked by simulating presentation of a large number of \q{pure} lists, i.e. lists consisting entirely of words with the same degree of polysemy k. The recall probability was averaged over all trials with the same value of k and the results plotted as a function of k. The three curves refer to lists of three different sizes, shown in the legend.
}
\centering 
\end{figure}
 
For all choice of the graph-structure parameters, the relationship between recall probability and the degree of polysemy of the word list is monotonously increasing. The more polysemic the words in the list, the easier each will be to recall. Rephrased in terms of word-length, this is nothing but the word-length effect, as exhibited by the diffusive model (Fig. 7). 
 
The reason for the word-length effect, in this model, is indeed a global or list-based mechanism: the fact that lists of shorter words, being more polysemic, produce a higher degree of interpretive clustering. 

When a word has a higher degree of polysemy, it takes on average a smaller distance to reach one of its meanings from anywhere within the semantic graph. Said otherwise, a diffusive particle will need to move less far if it has to interpret shorter words. For shorter words, therefore, the semantic region within which memories are formed will be narrower, and a smaller region will have to be explored during retrieval; thus, recall will be facilitated.   
  
This is shown in Fig. 8, where again distances on the page are meant to represent shortest-path distances within a denser semantic graph of which only a few nodes are shown. The nodes being shown refer to both some highly polysemic words (in shades of blue) and some highly oligosemic ones (in shades of red).  
Panel A of Fig. 8 shows the diffusive trajectory of the particle during the presentation of a list of polysemic words; panel B, during the presentation of a list of oligosemic words. In the latter case, the desired meanings are less readily available; so longer distances have to be travelled and, afterwards, the memories will have to be sought over a larger area of the graph. 
   
This is evidently not just an item-based effect. A comparatively long word, by causing a longer shift in the presentation trajectory, distances all the memories that will be created afterwards from the ones created before. Moving from memory to memory during the retrieval stage becomes, on average, harder over the full scale of the list size. 

\begin{figure}[h!]
\label{fig8}  
\includegraphics[width= .9 \textwidth]{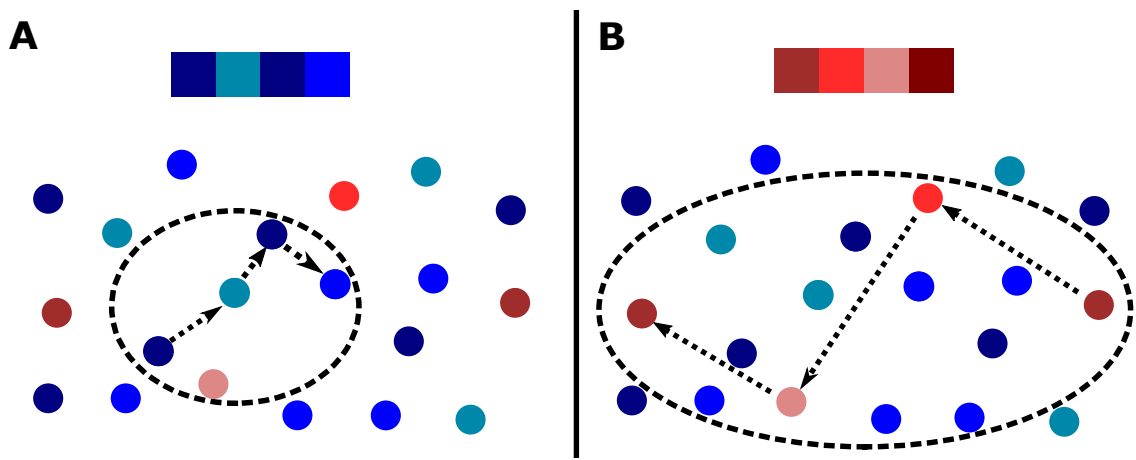}
\caption{Role of interpretive clustering in the word-length effect. Distances on the page are meant to represent shortest-path distances within a denser semantic graph of which only a few nodes are shown. These nodes refer to three highly polysemic words (shown in shades of blue) and three highly oligosemic ones (in shades of red). Dotted arrows depict diffusive motion through the semantic graph. Panel A depicts the diffusive trajectory during the presentation of a list of polysemic words; panel B, during the presentation of a list of oligosemic words, to the same system. In both panels, the list being presented is displayed over the drawing as a sequence of colored squares. In the oligosemic case, longer distances have to be travelled; therefore memories are distributed over a wider region (dashed ellipses), impairing recall. }
\centering 
\end{figure}

\section{Conclusions}
 
A diffusive-particle approach to the modeling of free recall has been developed, in which the presentation of words and their recall are modeled as trajectories of a particle diffusing over a semantic graph with a random structure. 
  
The model has predicted correctly some well-known features of free recall (forward asymmetry, the contiguity effect, the word-length effect). A novel prediction has also been obtained: shorter words, being more polysemic, are characterized by a stronger \q{sequentiality}, that is, they are more likely to be recalled through forward contiguity -- a prediction confirmed by a fresh analysis of archival data. 

The mechanism behind the latter phenomenon (\q{interpretive clustering}) is the same that lies at heart of the word-length effect as predicted by this theory. The conversion of words into meaning involves interpretation, and our freedom of interpretation (which is larger for the more polysemic words) has the effect of turning temporal contiguity into semantic contiguity. Since we memorize each word through a meaning largely determined by its context, mixed temporal-semantic correlations are created amongst memories. 
  
Future work on the theory may evolve in to three directions: (1) comparing results from simulations to additional features of available databases, or to features well-documented in the literature, some of which are not difficult to envision within this model (semantic contiguity effect, power-law scaling, recency); (2) trying out more realistic forms for the distribution $P(G)$ of the probabilistic graph through which the particle moves, and optimizing this distribution over the data; this may help in interpreting free-recall data as measurements of semantic connections within specific groups of words; (3) studying further the connections between this diffusive model and more widely tested retrieved-context models, in order to ascertain to what extent they differ and in what respects they may correspond.     

There are also several experiments that may help test the predictions made so far. In particular, it may be useful to perform ad-hoc experiments with select pools of words for which the measurement of polysemy is not overly tricky. This could be done by using two pools, one composed of decidedly oligosemic words (such as \q{Parthenon}) and one of extremely polysemic words (such as \q{let}).  
 
Experiments on such mixed lists would serve as a strict test for what we have claimed to be a polysemy effect in the sequential recall probabilities. Another task would be to test whether the word-length effect survives when each list harbors multiple word-lengths but is assembled entirely out of either pool (the highly polysemic or the highly oligosemic one). If recall probabilities do not depend on which pool has been used, that would disprove the explanation provided above, ruling out the role of interpretive clustering in the word-length effect. 
 
Finally, the degree of importance of interpretive clustering may be measured through experiments based on pseudowords. The meanings that a pseudoword inevitably evokes can affect its association value, playing a potentially important role in the recall process (Glaze, 1928); at the same time, the recall of pseudowords may be expected to be more largely phonetical than the recall of real words. If so, effects due to interpretive clustering will be reduced. Comparing data from experiments with words and from experiments with pseudowords may help ascertain to what extent semantics matters in the emergence of the effects we have discussed.

I would like to thank Michael Kahana, of the University of Pennsylvania, for generously sharing the data obtained in his laboratory. 

\section{Bibliography}
  
Baddeley, A.D., Thomson, N.,  Buchanan M. (1975). Word length and the structure of short-term memory. Journal of Verbal Learning and Verbal Behavior, 14, 575-589. 

Baddeley, A.D. (2007). Working memory, thought and action. Oxford: Oxford University Press.

Bousfield W.A. and Sedgewick C.H.W. (1944). An analysis of sequences of restricted associative responses. Journal of General Psychology (30), 149-165. 

Campoy, G. (2008). The effect of word length in short-term memory: Is rehearsal necessary? Quarterly Journal of Experimental Psychology, 61 (5) 724-734.

Cowan N. (2001). The magical number 4 in short-term memory: a reconsideration of mental storage capacity. Behavioral Brain Science, 24, 87–114. 

Ebbinghaus, H. (1913). Memory: A contribution to experimental psychology. New York, NY: Teachers College, Columbia University.

Farrell S. (2012). Temporal clustering and sequencing in short-term memory and episodic memory. Psychological Review, 119, 223–271. 

Frieze, A. and Karonski, M. (2016). Introduction to Random Graphs. Cambridge, UK: Cambridge University Press.

Fernando, C. (1996). Idioms and idiomaticity. Oxford, UK: Oxford University Press.

Glaze, J. A. (1928). The association value of non-sense syllables. Pedagogical Seminary and Journal of 
Genetic Psychology, 35, 255-269.

Greenberg, J. (1966). Universals of language. Cambridge, MA: MIT Press. 

Guiter, H. (1974). Les relations fr\'equence-longeur-sens des mots (langues Romaines et Anglais). In Atti del Congresso Internazionale di Linguistica (pp. 373-381). Amsterdam, Netherlands: Benjamins. 

Healey, M., Crutchley, P., Kahana, M.J. (2014). Individual differences in memory search and their relation to intelligence. Journal of Experimental Psychology, 143, 1553–1569

Howard, M. W., and Kahana, M. J. (2002). A distributed representation of temporal context. Journal of Mathematical Psychology, 46, 269-299. 

Hulme, C., Suprenant, A. M., Bireta, T. J., Stuart, G., and Neath, I. (2004). Abolishing the word-length eﬀect.  Journal of Experimental Psychology, 30, 98-106.

Katkov M., Romani S., Tsodyks M., (2014). Word length eﬀect in free recall of randomly assembled word lists. Frontiers of Computational Neuroscience (8), 129.

Jalbert, A., Neath, I., Bireta, T. J., and Surprenant, A. M. (2011). When does length cause the word length effect? Journal of Experimental Psychology, 37, 338-353. 

Jakobson R. (1987). Language in Literature. Ed. Krystyna Pomorska and Stephen Rudy. Cambridge, Massachusetts: Belknap Press. 

Kahana, M. J. (1996). Associative retrieval processes in free recall. Memory and Cognition, 24, 103-109.

Kahana, M. (2014). Foundations of human memory. Oxford, UK: Oxford University Press. 

Lohnas, L.J., Polyn, S.M., and Kahana, M.J. (2015). Expanding the scope of memory search: modeling intralist and interlist effects in free recall. Psychological Review, 122 (2), 337-363. 

Murdock B. (1962). The serial position effect of free recall. Journal of Experimental Psychology, 64 (2), 482-488.

Murray, D. J., Pye, C., Hockley, W. E. (1976). Standing's power function in long-term memory. Psychological Research, 38 (4), 319-331.

Musz, E.,  Thompson-Schill, S.L. (2015). Semantic variability predicts neural variability of object concepts. Neuropsychologia, 76, 41-51. 

Neath I., Bireta T.J., Surprenant A.M., (2003). The time-based word length effect and stimulus set speciﬁcity, Psychonomic Bulletin Review, 10 (2), 430-4

Nerlich, B., Todd, Z., Herman, V., Clarke, D.D. (2003). Polysemy: flexible patterns of meaning in mind and language. Berlin, Germany: Mouton de Gruyter.

Rensinghoff, S. and Nemcov\'a, E. (2010). On word length and polysemy in French. Glottotheory, 1, 83-88.

Romani, S., Pinkoviezky, I., Rubin, A., Tsodyks, M. (2013). Scaling laws of associative memory retrieval. Neural Computation, 25(10), 2523-2526.

Rothe, U. (1994). Wortlänge und Bedeutungsmenge: Eine Untersuchung zum Menzerathschen Gesetz an drei romanischen Sprachen. In R. Köhler and J. Boy (Eds.), Glottometrika 5, 101-112.

Sambor, J. (1984). Menzerath's law and the polysemy of words. Glottometrika 6, 94-114.

Standing L. (1973). Learning 10.000 pictures. Quarterly Journal of Experimental Psychology, 25, 207-222.

Xu, Z. and Li, B. Q. (2009). The Mechanism of Reverse Word Length Eﬀect of Chinese in Working Memory. 
Acta Psychologica Sinica, 41 (9), 802-811. 

Zipf, G. K. (1949). Human behaviour and the principle of least effort. Cambridge, MA: Addison-Wesley.

\end{document}